\documentclass[12pt,preprint]{emulateapj}
\usepackage{graphicx}
\usepackage{float}
\usepackage{amsmath}
\usepackage{amssymb}
\usepackage{enumerate}
\usepackage{verbatim}
\usepackage{multirow}

\newcommand{\ogwh}{\Omega_{\rm gw}h^2}

\begin{document}

\title{Projected constraints on the cosmic (super)string tension with future gravitational wave detection experiments}

\author{Sotirios A.~Sanidas, Richard A.~Battye and Benjamin W.~Stappers} \affil{Jodrell Bank Centre for Astrophysics, University of Manchester,
Manchester, M13 9PL, United Kingdom;\\ sotiris.sanidas@gmail.com, rbattye@jb.man.ac.uk, ben.stappers@manchester.ac.uk} \date{\today}

\begin{abstract} We present projected constraints on the cosmic string tension, $G\mu/c^2$, that could be achieved by future gravitational wave
detection experiments and express our results as
semi-analytic relations of the form $G\mu(\Omega_{\rm gw}h^2)/c^2$, to allow for direct computation of the tension constraints for future experiments. These results can be applied to new constraints on $\ogwh$ as they are imposed. Experiments operating in different frequency bands probe different parts of the gravitational wave spectrum of a cosmic
string network and are sensitive to different uncertainties in the underlying cosmic string model parameters. We compute the gravitational wave spectra of
cosmic string networks based on the one-scale model, covering all the parameter space accessed by each experiment which is strongly dependent on
the birth scale of loops relative to the horizon, $\alpha$. The upper limits on the string tension avoid any assumptions on the model parameters. We perform this
investigation for Pulsar Timing Array experiments of different durations as well as ground-based and space-borne interferometric detectors.

\end{abstract} \keywords{early universe --- gravitational waves --- instrumentation: miscellaneous --- methods: numerical} \maketitle

\section{INTRODUCTION}

Cosmic strings are one-dimensional topological defects of cosmological size, expected to form either during symmetry breaking phase transitions in the early Universe (field theory strings) \citep{kib76} or at the end of inflation in brane-world scenarios (cosmic superstrings) \citep{st02}.

The energy scale of cosmic strings is defined by their linear energy density or tension, $\mu$, usually referenced through the dimensionless quantity $G\mu/c^2$, where $G$ is Newton's constant and $c$ the
speed of light. In the case of field theory strings, the tension is directly related to the energy scale $\eta$ of the phase transition that created
them, $\mu\sim\eta^2$ in natural units, $\hbar=c=k=1$. In the case of cosmic superstrings, their tension is directly related to the fundamental string coupling
$g_{\rm s}$ and the compactification or warping scales. Therefore, the detection, or at least the constraining of the cosmic string tension,
provides a unique laboratory for physics at high energies.

A cosmic string network consists of infinite strings which stretch beyond our horizon and string loops \citep{vs94}. After its creation, the
network evolves along with the expansion of the Universe and is expected to settle into a scaling regime where all its fundamental properties scale
along with the horizon radius $\sim t$, which is called the one-scale model. The scaling evolution of the network is attained through the creation of loops. When two cosmic strings intersect, they intercommute with a probability $p$ and exchange partners,
chopping off loops from the network \citep{she87,jjp05}. In the one-scale model, loops are born with a size $\ell_{\rm b}$ equal to a fraction of
the particle horizon $d_{\rm H}$ at the time of their birth, $\ell_{\rm b}(t)=\alpha d_{\rm H}(t)$, where $\alpha$ is a constant. These loops subsequently
decay by emitting all of their energy in various forms of radiation, with the most dominant of them expected to be gravitational waves (GWs)
(\citet{vil81d}; see, \citet{vhs97} for an alternative). The GW emission from all the loops created by the network creates a stochastic
gravitational wave background (SGWB).

Recently, \citet{sbs12} have studied in detail the properties of the cosmic string SGWB and set a conservative upper limit on the cosmic string tension, $G\mu/c^2<5.3\times10^{-7}$ for networks with $\alpha\geq10^{-9}$, avoiding any assumption on the parameters describing the GW emission mechanism. Our result is approximately one order of
magnitude weaker than the equivalent limit found by the \citet{dv05} analytic relation, $G\mu/c^2<1.2\times10^{-8}$ \citep{vlj+11,vlj+12}, which is due to the strong assumptions made for the latter (see also discussion in \citet{svl12}).

As part of the common effort of the European Pulsar Timing Array (EPTA\footnote{http://www.epta.eu.org}, \citet{fvb+10}) collaboration to detect GWs with pulsar timing, we present here an expansion of our previous work to evaluate the tension constraints that will be achieved by future Pulsar Timing Array (PTA)
experiments of different durations and ground-based/space-borne interferometers. This not only allows us to predict future constraints, but the semi-analytic formulae we deduce can be used when the limits on $\ogwh$ become available. First, we establish the parameter space accessible by each
of these detectors. Then, we discuss how the exclusion curves on the parameter space evolve relative to $\Omega_{\rm gw}h^2$. Using this
information, we will evaluate the tension constraints as a function of $\ogwh$ for a range of experiments. We will also derive semi-analytic formulae for ease of use in future investigations. Additionally, we discuss how the radiative efficiency
parameter $\Gamma$, affects these
constraints.

\section{CONSTRAINING THE STRING TENSION}

\subsection{The cosmic string SGWB}

In the one-scale model, the basic parameters that determine the GW spectrum are: the cosmic string tension, $G\mu/c^2$; the birth scale of loops
relative to the horizon, $\alpha$; and the intercommutation probability, $p$. In order to compute the SGWB, we also need to model the spectrum of radiation emitted by a cosmic string loop. This is quantified by: its spectral index $q$, with $q=4/3$ for cusps and $q=2$ for kinks \citep{vs94}; a cut-off on the number of emission modes, $n_*=1\rightarrow\infty$, which is used to model the effects of gravitational backreaction \citep{cbs96}; and the radiative efficiency $\Gamma\approx50$, which describes the efficiency of the GW emission mechanism \citep{ca95}. In \citet{sbs12} we have described in
detail the computation of the cosmic string SGWB which we will only briefly recap it here.

From the energy-momentum tensor for cosmic strings we find the energy lost into loops in order for the network to maintain scaling,
\begin{equation}
\frac{dE_{\rm{loop,cr}}}{dt}=-V(t)\left[\dot{\rho}_{\infty}(t)+2\frac{\dot{a}(t)}{a(t)}\rho_{\infty}(t)\left(1+\langle\upsilon^2\rangle/c^2\right)\right]\,,
\label{eq::dEdt}
\end{equation}
where $\rho_{\infty}(t)=A\mu d_{\rm H}^{-2}(t)c^2$ is the energy density of infinite strings, $A$ is the number of
infinite strings within the horizon, $\langle\upsilon^2\rangle$ is the mean squared velocity of cosmic strings, $a(t)$ the scale factor and
$V=a^3(t)$ is the computation volume. We have considered $A$ and $\langle\upsilon^2\rangle$ as fixed parameters, using their most recently published values \citep{bos11}, quoted in Table~\ref{tab::params}. Assuming that loops have a length $\ell_{\rm b}(t)=\alpha d_{\rm H}(t)$ when they are born, from Eq.~\eqref{eq::dEdt} we
can find the loop formation rate,
\begin{equation}
\frac{dN_{\rm{loop}}}{dt}=\frac{2V(t)\rho_{\infty}(t)}{\mu \alpha
d_{\rm{H}}(t)c^2}\left[\frac{c}{d_{\rm{H}}(t)}-\frac{\dot{a}(t)\langle\upsilon^2\rangle}{a(t)c^2}\right]\,,
\label{dNdt2}
\end{equation}
where $N_{\rm loop}$ is the total number of loops within $V(t)$ created since the creation of the network at $t_{\rm f}=t_{\rm Pl}c^4/(G\mu)^2$, with
$t_{\rm Pl}$ the Planck time. Since the rate of decay of length for loops is $\Gamma G\mu/c$, we are able to compute the number density $n(\ell,t)d\ell$ of
loops with length between $\ell$ and $\ell+d\ell$ present at time $t$.

 In our GW emission model, each loop emits into a series of harmonics $n$, with frequency
\begin{equation}
f_{n}=\frac{2nc}{\ell}\,,
\label{eq::emfreq}
\end{equation}
and power $dE_{\rm{gw,loop}}/dt=P_nG\mu^2c$ with $P_n=\Gamma n^{-q}/\sum_{m=1}^{n_*}m^{-q}$. Then, using Eq.~\eqref{eq::emfreq} and carefully redshifting the frequencies to the present time we can construct $n_j(f,t)$ that
provides the number density of loops, which at the $j$th mode and at time $t$, emit GWs detected at frequency $f$. The computation of the energy density of GWs per logarithmic frequency interval $\ogwh$, where
$H_0=100h\,\rm{km\, s^{-1}\, Mpc^{-1}}$ is the Hubble parameter at the present time $t_0$, is
then given by
\begin{equation}
\Omega_{\rm{gw}}(f)=\frac{2G\mu^2c^2}{\rho_{\rm crit} a^5(t_0)
f}\sum_{j=1}^{n_{*}}jP_j\int_{t_{\rm f}}^{t_0}a^5(t^{\prime})n_{j}(f,t^{\prime})dt^{\prime}\,,
\label{eq::spectrum}
\end{equation}
where $\rho_{\rm crit}=3H_0^2c^2/8\pi G$ is the critical energy density.

A correction has to be applied in Eq.\eqref{eq::spectrum} due to the annihilation of massive particles during the radiation era. When the temperature of the Universe, $T$, drops below a specific particle mass threshold, the relevant family becomes non-relativistic, changing the number of relativistic degrees of
freedom, $g_*$. This will result in a faster expansion of the Universe every time such a mass threshold is crossed, affecting the value of $a(t)$ \citep{kt90a}. Although these changes will not significantly affect the evolution of the cosmic string network, which will quickly adapt to the new expansion rate and converge to its scaling evolution, they will have a significant effect on the energy density of the GWs emitted by the network.

\begin{figure}
\centering
\includegraphics[width=8.6cm]{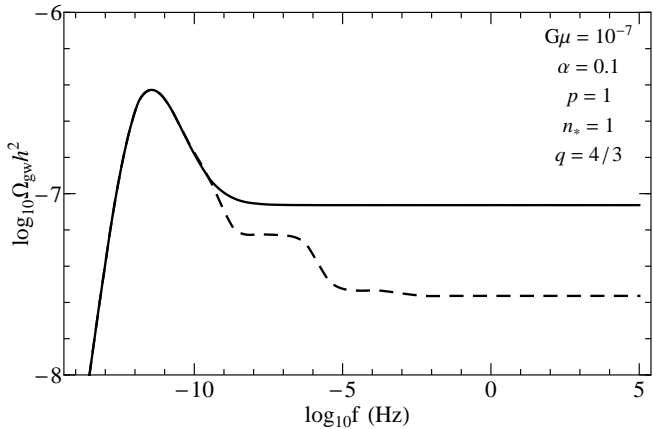}
\caption{The SGWB of a cosmic string network before (solid line) and after (dashed line) the
particle annihilation corrections are applied.}
\label{fig::spectrum}
\end{figure}

These effects can be included with the use of a multiplicative factor for $\rho_{\rm gw}$, and therefore $\ogwh$, which has the form $(g_*/g_*^\prime)^{1/3}$, where $g_*^\prime$ and $g_*$ are the
relativistic degrees of freedom before and after the transition, and acts to decrease the amplitude of the SGWB \citep{ben86b,ca92,bbcd12}. This correction is applied at
time $t_{\rm cor}=(32\pi G\rho/3)^{-1/2}$, where $\rho=\pi^2g_*T^4/30$ and its effects are observed at a frequency $f_{\rm
cor}\approx2a(t_{\rm cor})c[f_{\rm r}\alpha d_H(t_{\rm cor})a(t_0)]^{-1}$, which depends strongly on $\alpha$. PTAs, probing nHz GWs, are not significantly
affected by these corrections, and that only in the case where $\alpha$ is large. These corrections were not considered in \citet{sbs12}, but this does not affect the robustness of the results presented in the aforementioned paper. However, when considering GW detectors operating at the Hz and mHz bands significant corrections are expected. We have
computed these corrections for every particle in the Standard Model using the mass values in \citet{bab+12} and incorporated them to the computed
spectra. In Fig.~\ref{fig::spectrum} we present a typical cosmic string GW spectrum for $\alpha=0.1$ with the characteristic peak and flat part originating from
the loop populations in the matter and radiation eras respectively, along with the corrections to the spectrum due to particle annihilation.

\subsection{Exclusion Curves} \label{sec::excu}

The computation of the tension exclusion curves and upper bounds for a SGWB limit is done using the procedure described in \citet{sbs12}. First, we compute the GW spectra of cosmic string models covering all the accessible parameter space. Second, we find which configurations are compliant with the SGWB limit, both in terms of amplitude and assuming a flat local slope\footnote{Since we are only considering projected constraints by future experiments, an assumption for the spectral slope is necessary. We will discuss our choice later.}. Finally, we plot these configurations in the $G\mu-\alpha$ parameter space and the maximum tension allowed for a given $\ogwh$ limit will provide the tension upper limit.

 For a particular experiment, it is essential to determine the range of $\alpha$ values which are accessible to it. The SGWB of cosmic strings can span a wide range of frequencies, from below the nHz regime to beyond a THz, which depends on the size of the loops created by the network. The minimum frequency $f_{\rm min}$ of the SGWB is provided by the first emission mode of the loops which are born at $t_0$, and hence $\alpha_{\rm min}=f_{\rm min}f_{\rm r}d_{\rm H}(t_0)/2c$.

In this work, we will not treat the GW detectors as broadband, but instead we will designate each one a fiducial frequency; the one at which maximum sensitivity is achieved. For ground-based interferometers (hereafter,
``Hz-detectors'') and space-borne interferometers (hereafter, ``mHz-detectors''), we have assumed fiducial frequencies of $f=100\,{\rm
Hz}$ and $f=0.01\,{\rm Hz}$ respectively. For PTA experiments, the fiducial frequency is the inverse of the elapsed time over which observations were taken: $f=6.3\,{\rm nHz}$,
$f=3.2\,{\rm nHz}$ and $f=1.6\,{\rm nHz}$ for 5-, 10-, and 20-year PTA experiments respectively. They can only detect GW emission
for $\alpha\geq\alpha_{\rm min}$,
where $\log_{10}\alpha_{\rm min}=-9.5$ for a 5-year PTA, $\log_{10}\alpha_{\rm min}=-9.2$ for a
10-year PTA, $\log_{10}\alpha_{\rm min}=-8.9$ for a 20-year PTA, $\log_{10}\alpha_{\rm min}=-15.7$ for mHz-detectors and $\log_{10}\alpha_{\rm
min}=-19.7$ for Hz-detectors. Of course, real detectors are broadband and can observe at higher frequencies than the one that
maximum sensitivity is achieved, but, since they are typically only sensitive for a few orders of magnitude in frequency, $\alpha_{\rm min}$ provides a useful rule-of-thumb.

\begin{figure}
\centering
\includegraphics[width=7cm]{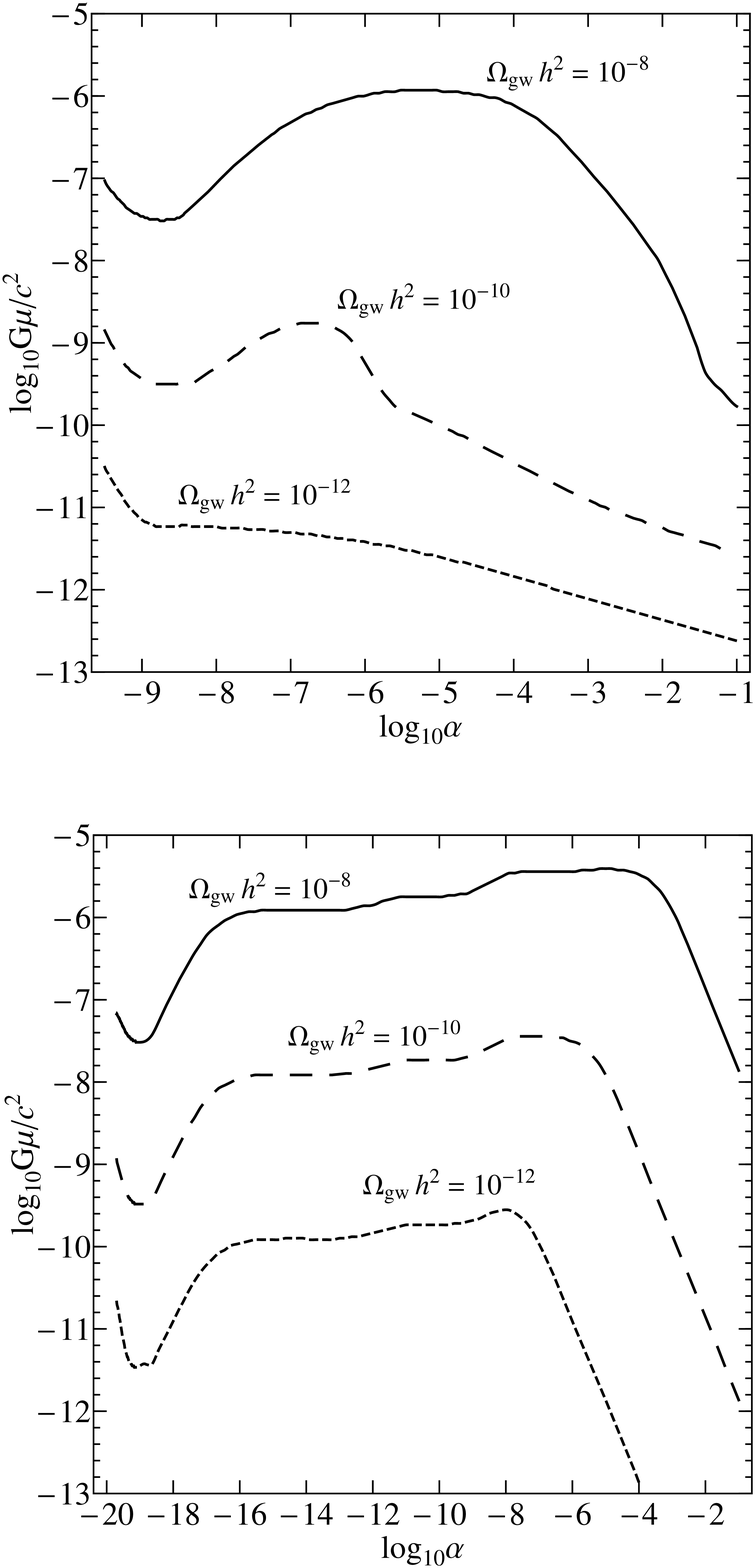}
\caption{Exclusion limits for cosmic string networks with $p=1$ and various values for the upper limit on
$\ogwh$ at $f=6.3\,{\rm nHz}$ (upper panel) and $f=100\,{\rm Hz}$ (lower panel).}
\label{fig::exclcurv}
\end{figure}

In Fig.~\ref{fig::exclcurv} we present the exclusion curves for networks with $p=1$ for three projected limits on $\ogwh$ at $f=6.3\,{\rm nHz}$ and
$f=100\,{\rm Hz}$. In the case of PTAs (upper panel), for $\ogwh\gtrsim10^{-10}$, the upper limit on $G\mu/c^2$ is provided by the peak in the mid-$\alpha$ region. The rise in the exclusion curve for $\alpha\sim\alpha_{\rm min}$ is expected, since in that region the detector probes the spectrum at frequencies lower than the peak frequency, with higher values of $G\mu/c^2$ required for a fixed $\ogwh$. For $\ogwh\ll10^{-10}$, the characteristic peak disappears and the upper limit on the tension is provided
by the models with $\alpha=\alpha_{\rm min}$.

The $\ogwh$ value at the low frequency cut-off is provided by the loops which are born at the present
time and therefore, it will not be stochastic. In reality, the SGWB is expected to start from a frequency $f\gtrsim f_{\rm min}$, where the individual loop emission becomes unresolvable, reducing the height of the secondary peak and overestimating the tension constraints for $\ogwh\lesssim10^{-10}$. However, since in that case the spectrum has a positive slope, our assumption for a flat spectrum leads to an underestimation of the tension constraints. These two effects are expected to be similar and cancel out, preserving the robustness of our results.

The shape of the exclusion curves for Hz- and mHz-detectors exhibit a much less prominent mid-$\alpha$ peak (see, for example, Fig.~\ref{fig::exclcurv}, lower panel). The whole mid-$\alpha$ region would be flat if we had not considered the massive particle annihilation corrections; a result of the fact that such detectors probe only the flat part of the spectrum originating from the radiation era. Moreover, the upper limits on the tension are always provided by the mid-$\alpha$ peak since the peak created by the networks with $\alpha\sim\alpha_{\rm min}$ never gets higher.

\section{PROJECTED CONSTRAINTS}

We computed the spectra for $\sim32000$ different parameter sets ($\sim1.3\times10^5$ CPU hours), using an updated
version of the code used in \citet{sbs12}. Most changes concern numerical accuracy improvements and the use of the recently published values for the
infinite string parameters from \citet{bos11}. A list of all these parameter values along
with the value ranges we covered for the fundamental cosmic string model parameters is presented in Table~\ref{tab::params}. It is only necessary to use a small number of spectral parameters, since the tension constraints for any
$\alpha$ in the $G\mu/c^2-\alpha$ parameter space will always be provided by the exclusion curves of either the $q=4/3,\,n_*=1$ or the
$q=4/3,\,n_*=10^4$ models as shown in \citet{sbs12}. Additionally, we only considered networks with $p=1$ since the final results can be easily rescaled for networks with
$p\ne1$. For the core results of this paper we used $\Gamma=50$ \citep{ca95}. \begin{deluxetable}{cc}[ht!] \tablecolumns{2} \tablewidth{0pc}
\tablecaption{Parameter values used/investigated in this work.} \tablehead{Quantity&Value(s)} \startdata $A$ (radiation era)&$45$\\ $A$ (matter
era)&$35$\\ $\langle\upsilon^2\rangle/c^2$ (radiation era)&$0.40$\\ $\langle\upsilon^2\rangle/c^2$ (matter era)&$0.35$\\ $f_{\rm r}$&$0.71$\\
$G\mu/c^2$&$[10^{-13},10^{-4}]$\\ $\alpha$&$[10^{-19.7},0.1]$\\ $q$&$-4/3\,,-2$\\ $n_*$&$1\,,10^4$\\ $\Gamma$&$50$,\,$[10,100]$\\ \enddata
\label{tab::params} \end{deluxetable}

\begin{figure*}
\includegraphics[width=17.5cm]{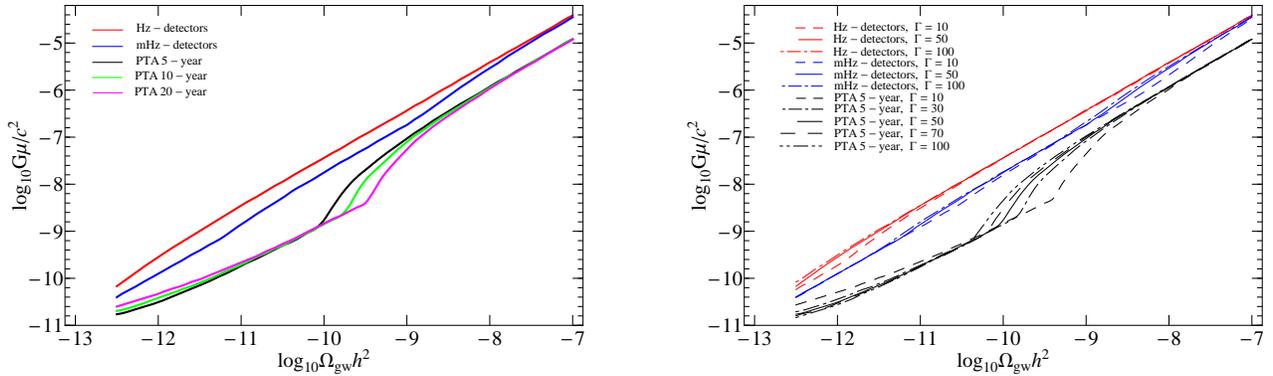}
\caption{Cosmic string tension upper limits for networks with $p=1$ as a function of $\ogwh$ for
various GW detection experiments (left panel) and the effects of varying $\Gamma$ (right panel).}
\label{fig::cons}
\end{figure*}

We have investigated SGWB amplitudes in the range $\ogwh\in[10^{-12.5},10^{-7}]$, with $\Delta\log_{10}(\ogwh)=0.1$. For each value for
$\ogwh$, we have computed the exclusion curves as a function of $\alpha$ and located the highest possible $G\mu/c^2$ value. In Fig.~\ref{fig::cons} (left panel) we present
the projected tension constraints as a function of $\ogwh$.

For PTAs, the constraints on the tension are almost independent of the frequency of maximum sensitivity except in the region $10^{-10}\lesssim\ogwh\lesssim10^{-9}$. Longer duration experiments are preferred in that region\footnote{Longer duration experiments will eventually collect more data and set more stringent constraints on $\ogwh$ than the shorter duration ones. However, if two such experiments manage to set the same constraint on $\ogwh$, the longer duration one will provide better constraints on the string tension in that region. The better performance of longer duration experiments stems from the lower frequency that they probe with maximum sensitivity, placing them ``deeper'' in the high amplitude, matter era peak of the cosmic string GW spectrum.}, which is the expected
sensitivity to the SGWB for LEAP \citep{ks10}. The change in the slope of the curve at $\ogwh\sim10^{-10}$ signifies the point at which the
tension upper limit is no longer provided by the mid-$\alpha$ peak but instead from the $\alpha\sim\alpha_{\rm min}$ models. The mHz-detectors will provide weaker constraints on the
tension for all $\ogwh$ values investigated. The same applies for Hz-detector, which will provide weaker constraints than both
PTAs and mHz-experiments.

The relatively simple origin of the constraint curves makes it possible to describe them with fitting formulae $G\mu(\ogwh)/c^2$. This will allow any
future constraint on $\ogwh$ by any experiment operating in the relevant frequency range to be directly connected to a robust constraint on the
string tension. Using least squares fitting and setting $x=\log_{10}(\ogwh)$, we find that:

 For 5-year PTA experiments,
\begin{equation} \log_{10}(G\mu/c^2)\leq\begin{cases} 0.0959x^3+2.292x^2+19.23x+50.31\\ {\rm for}\,-10<x\leq-7\vspace{0.5cm}\,,\\ 0.0917x^2+2.85x+10.5\\
{\rm for}\,-12.5\leq x\leq-10\,.\\ \end{cases} \label{eq::pta5} \end{equation}

 For 10-year PTA experiments,
 \begin{equation} \log_{10}(G\mu/c^2)=\begin{cases} 0.126x^3+2.972x^2+24.36x+63.22\\ {\rm for}\,-9.8< x\leq-7\vspace{0.5cm}\,,\\
    0.0794x^2+2.53x+8.52\\ {\rm for}\,-12.5\leq x\leq-9.8\,.\\ \end{cases} \label{eq::pta10} \end{equation}

 For 20-year PTA experiments,
 \begin{equation} \log_{10}(G\mu/c^2)=\begin{cases} 0.1565x^3+3.629x^2+29.05x+74.31\\ {\rm for}\,-9.5<x\leq-7\vspace{0.5cm}\,,\\
    0.074x^2+2.37x+7.46\\ {\rm for}\,-12.5\leq x\leq-9.5\,.\\ \end{cases} \label{eq::pta20} \end{equation}

 For mHz-detectors,
  \begin{equation} \log_{10}(G\mu/c^2)= 1.09x+3.16\,\,\,{\rm for}\,-12.5\leq x\leq-7\,. \label{eq::lisa} \end{equation}

 For Hz-detectors,
  \begin{equation} \log_{10}(G\mu/c^2)= 1.03x+2.83\,\,\,{\rm for}\,-12.5\leq x\leq-7\,. \label{eq::ligo} \end{equation}

These formulae can be easily adapted to describe cosmic string networks with $p\neq1$. In \citet{sbs12} we have pointed out that a non-zero
intercommutation probability only rescales the GW spectrum, $\ogwh\propto p^{-k}$, where $k$ is an index describing how the intercommutation
probability affects the population statistics of infinite strings, and whose value is discussed in \citet{sak05}
and \cite{as05,as06}. Therefore,
Eqs.~\eqref{eq::pta5},~\eqref{eq::pta10},~\eqref{eq::pta20},~\eqref{eq::lisa} and~\eqref{eq::ligo} can provide constraints for any possible
combination of $p$ simply by multiplying the polynomials with $p^{-k}$.

Additionally, we have investigated the effects of varying $\Gamma$ on the tension constraints. The value of $\Gamma$ depends on the trajectory of each loop and is expected to be in the range $\sim50-100$ \citep{ac94,ca95,boh11,cv11}. We have computed the tension constraints for networks with $\Gamma\in[10,100]$ and we present some typical results in the right panel of Fig.~\ref{fig::cons}. In the case of Hz- and mHz detectors the effects of varying $\Gamma$ do not significantly change the tension constraints. The same holds true for PTAs, except in the region $10^{-10.5}\lesssim \ogwh\lesssim 10^{-9}$ where low values of $\Gamma$ are equivalent to decreasing the frequency of maximum sensitivity.

We will conclude this section by discussing the consequences of our assumption of a locally flat spectrum. For Hz- and mHz-detectors, this
assumption holds true since these detectors probe the flat part of the spectrum for most of the $G\mu-\alpha$ parameter space. In the case of PTAs, the slope will be negative when $\ogwh\gtrsim10^{-10}$ (upper
limit provided by the mid-$\alpha$ peak) and positive when $\ogwh\lesssim10^{-10}$ (upper limit provided by the peak at $\alpha\sim\alpha_{\rm min}$). In
the first case, the flat spectrum assumption will lead to an underestimation of the constraint. For example, the EPTA limit on the string tension of \citet{sbs12} will provide a
hint on the expected discrepancy. Reworking that analysis under the assumption of a flat spectrum we find $G\mu/c^2<5.6\times10^{-7}$
whereas the limit we acquired using the extra slope information was $G\mu/c^2<5.3\times10^{-7}$, a $5.6\%$ discrepancy. We anticipate that the projected constraints
in the whole range where $\ogwh\gtrsim10^{-10}$ will be higher by a similar percentage, something that does not affect the robustness of our
approach since we are interested in conservative constraints. In the case where, $\ogwh\lesssim 10^{-10}$ the constraints are expected to be
overestimated by a similar percentage. However, expect that this will be canceled by a similar order underestimation of the constraints in this region due to the assumption of the stochastic nature of the background at $f_{\rm min}$, as discussed in Sec.~\ref{sec::excu}.

\section{DISCUSSION}

In this paper, we have presented fitting formulae for the upper limits on the cosmic string tension which can be used in future GW detection experiments. These constraints are free of any assumption on model parameters within the one-scale model. These formulae provide a conservative alternative to the approach discussed in \citet{dv05} which is
based on cusp emission and an approximate loop number density, assumptions which are much stronger.

The GW spectum of cosmic strings, being extremely broadband is potentially accessed by every GW detection experiment, with each of them probing
bands with different properties. PTAs, probing the high amplitude matter era peak of the spectrum for most cosmic string model parameter combinations and only partially suffering from the massive particle annihilation corrections in the case of large loops, provide an excellent tool to detect strings, or at least, set the most stringent constraints on their tension. On the other side, ground-based interferometers probe a much wider area of the parameter space ($\alpha_{\rm min}\approx10^{-20}$) making them the only tool able to detect emission from very small loops, and they are not sensitive to the, unknown, details of the radiation spectrum ($q$ and $n_*$). Space-borne interferometers, operating in a frequency range between PTAs and ground-based interferometers, share advantages and disadvantages from both.

PTAs already set the most stringent constraints on the
amplitude of the SGWB when compared to LIGO, and Advanced LIGO which is expected to become operational in 2015 will achieve a similar performance  \citep{har10}. The further increase in the sensitivity of PTAs
expected in the near future (LEAP is expected to set upper limits on $\ogwh\lesssim10^{-10}$), will lead to constraints on
$G\mu/c^2\lesssim 2\times10^{-9}$, with a SKA\footnote{http://www.skatelescope.org}-scale PTA experiment improving this by two orders of magnitude.

\section{ACKNOWLEDGEMENTS}

The authors would like to thank Michael Kramer, Mark Hindmarsh, and acknowledge the support of the colleagues in the EPTA during the preparation of this paper.

\end{document}